\documentclass[screen,sigconf]{acmart}
\AtBeginDocument{%
  \providecommand\BibTeX{{%
    \normalfont B\kern-0.5em{\scshape i\kern-0.25em b}\kern-0.8em\TeX}}}

\copyrightyear{2022}
\acmYear{2022}
\setcopyright{acmcopyright}\acmConference[CHI '22]{CHI Conference on Human Factors in Computing Systems}{April 29-May 5, 2022}{New Orleans, LA, USA}
\acmBooktitle{CHI Conference on Human Factors in Computing Systems (CHI '22), April 29-May 5, 2022, New Orleans, LA, USA}
\acmPrice{15.00}
\acmDOI{10.1145/3491102.3517628}
\acmISBN{978-1-4503-9157-3/22/04}

\begin{document}

\title{Coordination and Collaboration: How do Volunteer Moderators Work as a Team in Live Streaming Communities?}
% [Moderation Team Coordination and Collaboration in Live Streaming Communities]
% \affiliation{%

\author{Jie Cai}
\affiliation{%
  \institution{New Jersey Institute of Technology}
  \city{Newark}
  \country{USA}}
\email{jie.cai@njit.edu}

\author{Donghee Yvette Wohn}
\affiliation{%
  \institution{New Jersey Institute of Technology}
  \city{Newark}
  \country{USA}}
\email{yvettewohn@gmail.com}

\renewcommand{\shortauthors}{Cai and Donghee}

\begin{abstract}
Volunteer moderators (mods) play significant roles in developing moderation standards and dealing with harmful content in their micro-communities.  However, little work explores how volunteer mods work as a team. In line with prior work about understanding volunteer moderation, we interview 40 volunteer mods on Twitch --- a leading live streaming platform. We identify how mods collaborate on tasks (off-streaming coordination and preparation, in-stream real-time collaboration, and relationship building both off-stream and in-stream to reinforce collaboration) and how mods contribute to moderation standards (collaboratively working on the community rulebook and individually shaping community norms). We uncover how volunteer mods work as an effective team. We also discuss how the affordances of multi-modal communication and informality of volunteer moderation contribute to task collaboration, standards development, and mod's roles and responsibilities.
\end{abstract}

\begin{CCSXML}
<ccs2012>
   <concept>
       <concept_id>10003120.10003121.10011748</concept_id>
       <concept_desc>Human-centered computing~Empirical studies in HCI</concept_desc>
       <concept_significance>500</concept_significance>
       </concept>
 </ccs2012>
\end{CCSXML}

\ccsdesc[500]{Human-centered computing~Empirical studies in HCI}

\keywords{Teamwork; volunteer mods; content moderation; coordination; collaboration; rules and norms; live streaming; online community}


\maketitle
% \jie{Preprint accepted by CHI 2022}

\section{Introduction}
Harmful content, such as hate speech, online abuses, harassment, and cyberbullying,  proliferates across all different types of online communities. Live streaming is a promising and fast-growing industry. Many social media platforms have live streaming services (e.g., Facebook, Instagram, YouTube), even text-based online communities like Reddit have a live streaming subreddit. Reports show that all main live streaming platforms have big gains during the pandemic \cite{Droesch2021TheStatistics} and that the live streaming market is predicted to be worth over \$247 billion by 2027 \cite{Yanev202137+2021}. Live streaming as a novel type of online community provides ways for thousands of users (viewers) to entertain and engage with a broadcaster (streamer) in real-time in the chatroom \cite{Wohn2018ExplainingStreamers}. While the streamer has the camera on and the screen shared, tens of thousands of viewers are watching and messaging in real-time, resulting in concerns about harassment and cyberbullying to the streamer \cite{Uttarapong2021HarassmentNegativity, Zhou2021DesigningCyberbullying}. 
 
To keep a safe and civil online space, platforms and community administrators develop guidelines and enforce them with a combination of algorithmic approaches and human moderators (mods) (either paid or volunteer) to deal with harmful content, which is termed as \textit{``content moderation''}. Live streaming as a mixed media contains some unique attributes such as synchronicity and authenticity  \cite{Hamilton2014StreamingMedia}, making real-time moderation challenging. 

Given that many platforms heavily rely on human mods who are voluntary to support their communities like Twitch \cite{Seering2019ModeratorAlgorithms,Wohn2019VolunteerExperience}, it is essential to understand how these volunteer mods collaborate as a team, which has been by large overlooked with many HCI scholarships focusing only on the labor aspect \cite{Dosono2019ModerationCommunities}, mods' decision-making process \cite{Cai2021ModerationCommunities,Cai2021} and engagement  \cite{Seering2019ModeratorAlgorithms,Wohn2019VolunteerExperience}, and the moderation challenges \cite{Jiang2019ModerationDiscord}. Volunteer mods are not professional employees with good training and moderate content in an informal manner. While some studies more or less have mentioned that mods involve rules development and communication with other mods, the discussion about human mods usually treats mods as individual entities. There is a lack of a specific piece to understand mods' coordination and collaboration, a gap this work intends to fill. Exploring moderation team coordination and collaboration can potentially help avoid team conflicts, improve team relationships, and consequently, help build an effective team to grow the community.

In line with recent work to unpack the opaque operation of content moderation (e.g., \cite{Steiger2021TheSupport,Jhaver2019DoesReddit, Wohn2019VolunteerExperience}), we interview 40 volunteer mods in live streaming communities to explore how they contribute to the moderation standards and collaborate on tasks as a team. Our work mainly contributes to understanding the team collaboration of volunteer mods in the moderation context. We reveal how the streamer and mods work as an effective team and the nuanced differences of volunteer moderation in live streaming communities. Our findings can benefit multi-stakeholders, such as the community admin, each volunteer mod, and the admin and mods as a team for the community growth.

\section{Research Context: Live Streaming Platform Twitch}

Twitch is the leading live streaming platform and consists of thousands of streaming channels centered on streamers. Reports show that 3.8 million streamers were broadcasting on Twitch in 2020 \cite{Yanev202137+2021}. Each streamer owns the power to assign or revoke other users as mods with special badges to indicate their mod status. When the streamer goes live, each viewer can clearly see how many viewers are watching at the time and check the viewer list to know other viewers' usernames. Each viewer can also see the active mods online at the top of the viewer list. Most viewers apply pseudonymous usernames without showing much information about their identities. Viewers can send messages in the public chatroom or send a Whisper message to a viewer privately by clicking on the viewer's username. Streamers can post their streaming schedules to their homepages so that viewers can know in advance. Followed or subscribed viewers can also receive email notifications before the streaming.  

Twitch applies the combination of human mods and moderation tools to govern online communities. At the platform level, it has community guidelines and terms of service, which are enforced by paid staff to mainly deal with video content violation and user reporting. At the channel or micro-community level, it allows streamers and their appointed mods to set up channel rules and even specific chat rules. Mods are selected from the streamers' close friends or active viewers in the community \cite{Wohn2019VolunteerExperience}. In some micro-communities, the experienced mods can be the head mods. Many micro-communities consist of only the streamer and other mods without a clear hierarchy. A streamer can also be considered a mod with full access to moderation functions. To support moderation, Twitch also provides a moderation tool called ``AutoMod'' to facilitate mods to filter certain words and punish violators with commands (ban, timeout, delete); Twitch also provides open API to third-party developers to develop bots with more customized features to support the team to conduct moderation tasks \cite{Cai2019CategorizingTwitch}. 

We chose Twitch as the research context because it is a leading platform in a promising industry, but there is limited understanding about volunteer moderation teams. In this study, we focus on volunteer mods led by the streamer to form a moderation team and explore how they work together to moderate the chat messages and viewers in an interactive environment. 

\section{Related Work}

\subsection{Volunteer Moderation and Remote Collaboration}
\subsubsection{Content Moderation, Volunteer Moderators, and Moderation Practices}
Content moderation refers to \textit{``the governance mechanisms that structure participation in a community to facilitate cooperation and prevent abuse''} \cite{Grimmelmann2015} and is achieved by the collaboration of human mods and algorithms for most platforms. Algorithmic content moderation can deal with obvious violations at scale but lacks of context-sensitivity, as Gillespie \cite{Gillespie2020ContentScale} points out that \textit{``automated content moderation is not a panacea for the ills of social media, and maybe contrary to the principles of governance that platforms should be pursuing.''} Human mods are still heavily relied on by most platforms, either paid laborers hired by or contracted with the platform (commercial mods) \cite{Roberts2016CommercialWork,Gillespie2018CustodiansMedia}, or free laborers with voluntary participation to help the community (volunteer mods) \cite{Wohn2019VolunteerExperience}. 

Prior work in HCI and CSCW about content moderation has broadly discussed the physical and emotional labor \cite{Roberts2016CommercialWork,Dosono2019ModerationCommunities}, psychological well-being \cite{Steiger2021TheSupport}, rules and norms \cite{Chancellor2018NormsCommunities}, moderation mechanisms such as flag, removal, and ban \cite{Kou2021FlagModeration, chandrasekharan2017bag, Srinivasan2019ContentCommunity}, relation with either the end-users or the admins \cite{Wohn2019VolunteerExperience, Cai2021}, and the support with various design interventions (see meta-analysis by \cite{ Feng2020CategorizingInterventions}).

In many online communities consisting of thousands of micro-communities (e.g., Reddit with subreddits, Discord with servers, and Twitch with streamer channels), volunteer mods play a major role in governing their micro-communities. A group of research involving volunteer mods often treats them as individual entities to conduct the moderation practices, such as the moderation tools they have used \cite{Cai2019CategorizingTwitch,Jhaver2019Human-machineAutomoderator} and the different roles they have played in the community \cite{Seering2022MetaphorsModeration, Wohn2019VolunteerExperience, Park2012FacilitativeERulemaking}. These roles involve various tasks and moderation strategies, such as helping the streamer to manage viewers \cite{Wohn2020AudienceTwitch} and communicating with violators  \cite{Cai2021ModerationCommunities}.

 The application of moderation strategies and doing tasks requires much communication and coordination. Though some work points out the discussion among other mods or the admin (e.g., \cite{Cai2021ModerationCommunities,Seering2019ModeratorAlgorithms}), there is a lack of work to investigate mods collaboration and coordination mechanism. This work extends existent research in volunteer moderation regarding mods' individual decision-making to the collective decision-making process for two reasons. First, compared with commercial mods working in the formal environment with training and clear instructions to follow \cite{Gillespie2020ExpandingDebates}, volunteer mods might need more collaboration and coordination to figure out everything on their own because of the informal environment with vague moderation guidelines, consequently taking more of the decision-making roles in the process. Second, since volunteer moderation often involves many users in the governance process \cite{Grimmelmann2015},  it is essential to understand how these users work as a group because many online communities need to break into micro-communities as they grow.

\subsubsection{
Remote Collaboration and Teamwork}

Existent research about remote collaboration is well-grounded from offline to online in both HCI and CSCW. In an offline context, remote collaboration is well-established from empirical and theoretical perspectives. For example, lots of work has explored paradigms and prototypes to overcome the shortcomings of remote work in decades, such as distance and physical presence (e.g., \cite{Ishii1997TangibleAtoms,Elrod1992Liveboard:Collaboration,Kuzuoka1992SpatialCapability}). In socio-technological conditions, Olson and Olson \cite{Olson2000DistanceMatters} propose that four key components determine effective remote collaboration, also called \textit{``distance framework''} \cite{Bjrn2014DoesCollaboration}: common ground, coupling/dependency of work, collaboration readiness, and collaboration technology readiness, and suggest that groups, with high common ground and loosely coupled work, with readiness both for collaboration and collaboration technology, have the potential to succeed with teleworking, and that deviations from each component restrict teammates' performance and require changes during the collaboration.

Collaboration is a high order type of collective action and has higher interaction, intensity, and integration compared with coordination \mbox{\cite{Keast2007GettingStrategies, Whelan2017ManagingCollaboration}}. Coordination is a process of integrating and aligning the actions, knowledge, and objectives of group members to achieve common goals \cite{Freeman2019UnderstandingCoordination, Rico2008TeamApproach}. There are two types of coordination: explicit coordination requiring communication to coordinate group activities and implicit coordination requiring shared mental models and common grounds to avoid overt communications \cite{MacMillan2004CommunicationCognition., Freeman2019UnderstandingCoordination}. In this work, we disentangle coordination and collaboration; we use collaboration to describe practices involving much communication; we use coordination to describe practices involving more guidance and direction and less interaction.

In online collaboration, there are two main threads of research in HCI and CSCW. One thread is about distributed teamwork with organized, stable, and long-last collaboration in contexts such as teleworking, software development, and education (e.g., \cite{Bjrn2014DoesCollaboration,Bjrn2009VirtualTranslucence,Eubanks2016TeamFulfillment,Hossain2004ICTTrust,Sere2011OnlinePerformance}); another is about virtual teamwork with informal, unstable, and short-term collaboration in contexts such as gaming (e.g., \cite{Dabbish2012CommunicationTeam, Leavitt2016PingGames,Freeman2019UnderstandingCoordination}) and peer production communities (e.g., \cite{Kittur2007HeWikipedia,Kittur2008HarnessingWikipedia,Kittur2010BeyondGroups}). 
In the informal, unstable, and short-term collaborative process, many scholars have investigated the group characteristics of the volunteer workers and how these characteristics are related to group productivity and outcome quality (see meta-analysis by \cite{Arazy2011InformationConflict,Mesgari2015TheWikipedia}). For example, the interest and experience diversity of the editor team on Wikipedia can positively influence the quality of articles \cite{Sydow2017DiversityWikipedia}; teams should maintain a balance between administrators and content creators as both contribute to the collaborative process \cite{Arazy2011InformationConflict}. As the community grows, the conflicts increase, and costs of coordination, such as conflict resolution, consensus building, and community management, also increase \cite{Kittur2007HeWikipedia}. Coordination mechanisms were not always effective for managing different conflicts in the team \cite{Kittur2010BeyondGroups}, and both implicit coordination through editor concentration and explicit coordination through communication can potentially improve the article quality when they were used \cite{Kittur2008HarnessingWikipedia}. While most scholarships focus on the editorial team and the visible outcome -- though some have mentioned that editors can revert vandalism, little work explores the moderation team and the invisible content management, namely, how volunteer mods work as a team to get rid of harmful content, such as vandalism on Wikipedia and harassment on Twitch.

While remote collaboration is well-established in contexts targeting professional workers and unprofessional volunteers in peer production communities, it is less clear how unprofessional volunteers collaborate in the content moderation system, usually lacking transparency \cite{Jhaver2019DidReddit,Roberts2016CommercialWork, MyersWest2018}. 
These mods are voluntary to work behind the scenes to deal with harmful content. However, little is known about how they collaborate on tasks. The synchronicity of live streaming might also cause the team to work differently compared with asynchronous communities, because the real-time affordance brings all mods online simultaneously, and mods have to make immediate decisions as the chat messages flow. The synchronicity also indicates that a single mod can't effectively concentrate on much information generated in real-time; working as a team with instant and continuous monitoring is imperative. Thus, we ask: 

\begin{itemize}
\item \textbf{RQ1}: How do mods collaborate on moderation tasks in the moderation team?
\end{itemize}

\subsection{Community Norms and  Rules}

Norms are \textit{``regularities in attitudes and behavior that characterize a social group''} \cite{Hogg2006SocialNorms}. Norms play significant roles in shaping online communities by indicating the group identity and regulating user behaviors \cite{Cialdini1991ABehavior,Chancellor2018NormsCommunities}.  
Online community norms have been well-established over the past few decades and have not halted as online spaces evolve and diversify (e.g., \cite{Cialdini1991ABehavior,Blanchard2010ACommunity,Larsson2012LawCommunity,Markman2012ASites,Zhou2011UnderstandingPerspective}).

In online communities applying volunteer moderation, norms vary across micro-communities, different from guidelines and terms of service aggregated by the platform and applied to all stakeholders. Much work has explored norms' impact on user's perceptions and behaviors, such as setting a good interaction example \cite{Seering2017ShapingExample-setting} and clarifying norms \cite{Cai2021ModerationCommunities,Jhaver2019DoesReddit} to mitigate harassment behaviors.

Norms are different from but interrelated with rules and can be transferred regarding the explicitness \cite{Cialdini1998SocialCompliance.}. Some work does not distinguish rules and norms and uses them interchangeably (e.g., \cite{Tekinbas2021DesigningMinecraft, Gilbert2020iR/AskHistorians}); other work applies the formalization to disentangle rules and norms like formalized rules and informal norms \cite{Fiesler2018RedditGovernance,Chandrasekharan2018TheScales}. Overall, norms can be finalized into well-written rules and guidelines. Some work focuses explicitly on the rules and tries to understand rule content and expression \cite{Fiesler2018RedditGovernance,Cai2021UnderstandingTwitch}.

Norms are usually about the user's perception of other users' thoughts and about the identity and behavioral conformity with other users in the group \cite{Cialdini1998SocialCompliance.}. We focus on the mods' perception of rules and norms they follow. In this work, we use \textit{``moderation standard''} to indicate what mods follow to conduct moderation, including the explicitly written text of rules (e.g., chat rules, Twitch community guidelines, terms of service) and the implicit norms among the moderation team or in the micro-community. Much prior work explores the \textit{``what''} question about rules and highlights the significance of rules and norms for community development (e.g., \cite{Chandrasekharan2018TheScales, Chancellor2018NormsCommunities}). Some have mentioned that the mods can help the admin to develop rules for the community \cite{Seering2019ModeratorAlgorithms}, but lack specificity in the nuanced context in volunteer moderation and clear answers to the \textit{``how''} question. Thus, we ask:

\begin{itemize}
\item \textbf{RQ2}: How do mods contribute to community moderation standards?
\end{itemize}
\section{Methods}
\subsection{Participants Recruitment}
This project was approved by the school's Institutional Review Board (IRB).  We recruited 40 volunteer mods on Twitch in many ways.  First, we used our lab's Twitter accounts to post recruiting messages and also to reach Twitter users who were tagged as \textit{``Twitch mod'' }and \textit{``moderator on Twitch''}.  We obtained ten mods through Twitter.  Second, we directly contacted active mods on Twitch.  The authors and the research assistants used their Twitch accounts and randomly browsed the recommended channels on the Twitch homepage and other channels that had many active mods.  Taking advantage of the Twitch Whisper function, we directly sent recruiting messages to each mod and asked them to refer to their friends if they were not interested.  We obtained 12 mods through Twitch Whisper.  Third, we used the email list that we collected from Twitch Convention 2019 and obtained six mods.  Fourth, we asked the Twitch streamers who had participated in our research to recommend mods and obtained five mods.  Lastly, two research assistants who were mods on Twitch used their networks to obtain seven mods.  We intended to keep the sample diverse, so we searched for participants with diverse experiences such as moderation years and categories.  Demographics were detailed in \autoref{tab:demo}.  The average moderation tenure was around three years (\textit{M}= 2.74, \textit{SD}= 1.63), ranging from half a year to eight years.  The average age was around 27 (\textit{M}= 26.51, \textit{SD}= 8.20). Most mods were White (65\%), followed by Asian (12.5\%), Africa American (10\%), Hispanic (5\%), and Pacific Islander (2.5\%).  Most were male (67.5\%), followed by female (27.5\%), transgender male (2.5\%), and transgender female(2.5\%).  The main streaming content was gaming.

\subsection{Interviews and Analysis}
We conducted semi-structured interviews through Discord (a VoIP communication application that is often used by Twitch streamers and mods).  In the interview, we first asked some general questions about their moderation experience and the content categories, such as \textit{``Who are you a mod for?''} and \textit{``How long have they been a mod for?''}.  Then we asked some questions related to our research questions, such as 1) How do you decide what is appropriate or not?  2) Who comes up with the criteria?  With follow-up questions about how do you/ they come up with moderation criteria?  3)Do you communicate with other moderators?  How?  Why?  4) How do you coordinate (during stream) if multiple people are modding at the same time?  In the end, we asked for their demographic information such as age, race, and gender.  All interviews were audio-taped, transcribed by speech recognition software, and then reviewed by researchers.

We imported all transcriptions into ATLAS.ti \footnote{https://atlasti.com/} for open coding. we used thematic analysis \cite{Braun2006UsingPsychology} to code answers into concepts and group the relevant concepts into themes.  First, two researchers went through each transcript and the interview protocol to overview the questions and answers; consequently,  they highlighted questions related to our research questions.  Second, they conducted the open coding for the first two transcripts individually.  In a weekly meeting, they discussed and clarified their codes to reach a mutual agreement about the coding process.  Next, one researcher completed the coding following the criteria developed through the discussion in several weeks, each week with a calibration meeting to present and clarify codes.  The coding process lasted for approximately one month.  Finally, two researchers exported the 117 codes into the Miro \footnote{https://miro.com/} whiteboard to collaboratively and iteratively organize codes into groups, categories, and high-level themes.

\begin{table*}
  \caption{Participant Table}
  \label{tab:demo}
  \resizebox{\textwidth}{!}{\begin{tabular}{@{}cccccc||cccccc@{}} 
 % \begin{tabular}{ccccccc||ccccccc}
    \toprule
    ID & Mod (yrs)& Age & Race & Gender & Content & ID & Mod (yrs)& Age & Race & Gender & Content \\
    \midrule
P1&	2&	18&	White&	M&	gaming&		P21&	1&	33&	White&	F&	gaming, creative\\
P2&	-&	-&	White&	M&	-&		P22&	1.5&	24	&White&	M&	gaming	\\
P3&	1.5	&37&	Asian&	F&	board games& P23&	6&	31&	White&	M&	gaming, chatting IRL, e-sports\\
P4&	2&	35&	White &	M&	gaming&		P24&	2.5&23&	Hispanic&	M&	art, body painting\\
P5&	2.5&41&	White&	M&	music, creative& P25&	1.5&	18&	White&	M&	gaming\\
P6&	1&	29&	White&	M&	-&	P26&	3&	27&	African American &	F&	gaming\\
P7&	2&	19&	White&	M&	gaming&		P27&	2&	21	&Asian&	M&	gaming, chatting IRL\\
P8&	2&	40&	White&	F&	gaming&		P28&	4&	18&	White&	F&	gaming, video editing\\
P9&	3-4&40&	White&	M&	gaming&		P29	&3	&29&	White &	M&	gaming, life advice, politics, drama\\
P10&1&45&	Africa American	&Trans male&	gaming&		P30	&4	&21&	Hispanic&	F&	gaming\\
P11&2-2.5&	23&	Asian&	M&	gaming	&P31	&3.5&	34&	White&	F&	gaming\\
P12&5&	27&	Africa American &	F&	gaming& P32 &	3&	19&	White&	F&	gaming\\
P13&1&	20&	White&	M&	gaming&		P33&	1&	19&	White&	M&	gaming\\
P14&1&	21&	-&	M&	gaming&		P34	&5&	26&	Asian&	M&	gaming\\
P15&4&	-&	-&	M&	gaming&		P35&	3&	24&	Pacific Islander&	M&	gaming, car racing	\\
P16&4&	24&	White&	M&	-&		P36&	1&	20&	White&	F&	gaming	\\
P17&3&	21&	White&	M&	gaming&		P37	&4&	19&	African American& 	M&	gaming\\
P18&2&	-&	White&	Trans female&	board games	&P38&	2&	18	& Asian	&M&	gaming\\
P19&5&31&	White&	M&	-&		P39&	0.5	&15&	White&	M&	rhythm \& music game	\\
P20&-&	43&	White&	M&	gaming, product reviewing&	P40&	8&	28	&White&	F&	gaming, chatting IRL, politics\\

\end{tabular}
}
\end{table*}

\section{Results}

\subsection{Ways to Collaborate on Tasks}
The first research question inquired how mods collaborated on moderation tasks in the team. To make sure there were enough active mods who could actually work with the team when the streamer was online, the team used many ways to collaborate: coordinating and preparing off-stream, collaborating in real-time in the stream, and fostering relationships both off-stream and in-stream to reinforce collaboration. Though many mods expressed that as individuals, the tasks had no fixed schedule and were random, they were on call as a team.

\subsubsection{Coordinate and Prepare Off-stream}

Off the stream, the streamer and other mods purposely recruited new mods from different time zones to cover the streamer's streaming slot; the streamer or the head mods also notified the team to prepare for streaming events. All these methods were to ensure enough active mods online in the stream.

The moderation tasks faced many challenges. Some channels appointed many mods but failed to have enough active mods. P36 said that all mods had different schedules, such as jobs and schools, and no one would like to moderate at night for a long time. Similarly,  P1 (M, 18) stated, \textit{``There have been some incidents where it's been 3 am and I am being in chat, and I'm the only mod in chat. Literally, I have to wait until I see another mod show up before I can actually go to bed.''} If mods worked for multiple channels at the same time, they would like to prioritize their preferences. For example,  P36 (F, 20) complained, \textit{``We got some new mods, but most of them are moderators for other channels. So they prioritize those. So sometimes it's really like a struggle to find moderators that are active. ''} According to P36, though you have many active mods, they might not engage in moderation in the expected channel. Additionally, too many mods reduced the workload but also caused \textit{``competition between mods''} (P21, F, 33). Many mods said that if there were two to three active mods in the chat, they would self- deactivate to work on their own things.

\paragraph{Purposely Recruit Mods from Different Time Zones}

Since individual mods had different schedules in their offline life and failed to cover the streaming slots all the time, the moderation team would like to recruit mods from different time zones to \textit{``make sure there's plenty of coverage and availability''} (P31, F, 34). 

\begin{quote}
Usually, whenever for the large chats when they wanna assign shifts and want 2 moderators at every time, what they'll end up doing they'll have a schedule, and you have to be at this time, this time, and this time. They'll end up taking 3 mods from CA, so they're on US west. We'll have 3 mods from NY, so they're on the US east and 3 mods from Europe; they just have 3 mods from every time zone they can find... but for smaller channels, it's not really a big deal because the streamer themselves could moderate the chat on their own. It's just whenever someone shows up and is willing to help; then it's beautiful. (P15, n/a)
\end{quote}

P15 showed a good example for a large stream with a clear schedule to have mods in different time zones to cover the streaming slots. He also pointed out that for small streams, the streamer might work as a mod to handle the chat in a comparatively flexible manner. 

\paragraph{Notify Each Other of the Attendance  Before the Streaming}

Streamers as the leaders in the moderation team would notify mods and schedule attendance for particular events with the expectation to have a lot of viewers (e.g., Twitch front-page streaming, anniversary stream) or with the prediction of an amount of moderation work (e.g., a restart of streaming after the ban, streaming new content). This method was usually applied to large streams with big events.  

Several mods mentioned that they coordinated to prepare for events. The streamer might \textit{``need all hands on deck''} and asked mods \textit{``as many as possible to be there at this time.''} (P19, M, 31). For example, P31 (F, 34) shared her experience, \textit{``Recently Z got told certain dates and times that his streams will be on the front page of Twitch, and he was like, Hey, just so you know, we're probably gonna have more people for these days. I'm letting you all know ahead of time.''} 

\begin{quote}
When there's a big thing going on, like when people get put on the front page of Twitch, when the streamer we moderate for does that, we definitely talk about who can be there because it's a lot of people. The front page of Twitch can bring in a lot of people. So, we need to be able to reference who can be there, and we need at least three or four people there. So yeah, we do sometimes when big events are happening for sure. (P28, F, 18)
\end{quote}

In these cases,  streamers needed to coordinate with the stream to ensure \textit{``at least at least three or four''}, \textit{``more''}, even \textit{``as many as possible''} mods online.  Not informing mods before events made moderation work overwhelming and caused discomfort and complaint from mods (P18, Trans, n/a).  

If the streamer predicted that incoming streaming needed more content moderation than regular streaming, they might also inform the team to ensure there were enough active mods online. For example,  P34 stated that the streamer planned to stream new content with no idea about their performance and viewers' reaction, so they scheduled a time to ensure mods were online, similar logic in another case reported by P24 (M, 23): \textit{``That streamer that got banned and she says this is going to be my first night back and I'm going to have a huge target on my back and people are going to be scrutinizing everything I do. I need you guys here. So we'll actually sit down and say we need to decide who's going to show up, who can make it, and what are we going to do.''} According to P24,  the streamer restarted streaming and predicted to have a lot of harassment in the chat, so that they \textit{``sit down''} to discuss which mods should \textit{``show up''}.

Mods also informed streamers if they couldn't attend events or streaming. P21 (F, 33) shared her experience to inform the team of her availability so the team could prepare in advance: \textit{``I'm saying if I went to my parents or something, I would let the streamer know that I'm not going to be there for the week and I would also say in the mod chat, so they'd be aware and so somebody else might for that stuff.''} P23 (M, 31) further explained, \textit{``If you're part of a moderation team, you gotta let people know so that they know what to expect from you. If you set appropriate expectations, people are going to be fine 99\% of the time.''}

\paragraph{Update Information in the Team}

Mods also updated each other off the stream about the streaming or chat in general. P24 (M, 23) would like to have mod meetings to update everything: \textit{``I try and at least do it every two weeks because people are busy. Right? It's difficult to make time sometimes cause they're busy. So I tell them it's even if you can't make the meeting, you know, to at least take some notes, and you know, look at what we can do going forward.''}  P24 suggested mods take notes to keep updated. P30 (F, 21)  coordinated with other mods to keep each other updated: \textit{``Just to keep in check with what's going on; like if I miss a day, keep them update me on what's happened; or if he misses a day, I'll update him on what's happening. It's just always to be clear of  what's happening in chat and in the stream.''} These updates usually happened in Discord with text messages so that everyone in the team could see them. \textit{``If somebody misses something, they can always go there and catch up on it.''} (P23, M, 31).  

\subsubsection{Collaborate in Real--time In the Stream}
In the stream, mods and the streamer involved a lot of communication via mainly third-party platforms like Discord and Skype. They also used the Whisper chat on Twitch for small group discussion and immediate communication. P28 (F, 18) said that they were all in the moderation discord for mods only and \textit{``talk in there when things are happening or when we need to get to know each other.''} Most moderation teams preferred Discord as the main communication channel because it could easily reach out to the whole team, was the main space to socialize off-stream, and archived all important information.

\paragraph{Monitor and Inform the Team Regarding Potential Incidents}

Mods shared information in the team so each team member was well-informed and notified during the moderation process, such as a shared list of repeated violators, a full list of information that the team could refer to in specific situations, and a notification of specific activities and violators in the Discord. Some mods notified the team to monitor specific activities or viewers. For example, 

\begin{quote}
We're all on the same page, so somebody will get to it first before someone else, it's fine. If somebody bans somebody or if somebody times out somebody out that's fine, we trust each other. If there's anything strange going on we will talk in the moderator chat in the Discord and let people know, ``hey this person is saying kind of funny things, let's watch them,'' or sometimes we'll DM each other, but usually the mod chat is better because all the mods and the streamer can see it. (P10, Trans, 45)
\end{quote}

P10 stated that mods in the team trusted each other's moderation actions; they notified the team on Discord about the suspicious activities in the streaming chat and also preferred the Discord chat to the Twitch chat because all team members could see it. Some mods also recognized viewers from other streams and notified mods in the current stream: \textit{``Since I mod for a lot of channels, I might recognize a viewer from a different stream. So I'll go over to the mod channel and say, ‘hey, that viewer, he caused a problem in another stream, so you guys know about that' ''} (P5, M, 41). When they saw these viewers who might potentially target the community, they \textit{``just head up, look out for these people''} (P27, M, 21).

The moderation team also collaboratively tracked violation activities with violators' information. Mods kept working on lists of violators and records of violation and used these pieces of information to supplement their monitoring. P23 (M, 31) said, \textit{``In that community, in particular, we have a shared Google document. It has a list of, um, I'm not finding the word, but people that have caused problems in the past for timeout or bans or whatever. If I lack context in a situation in that community, then I can go to that spreadsheet.''} Similarly, 

\begin{quote}
    Well, first off, you know, we monitor as a group. Another thing, I have our Discord open while the stream is there, and if it's something more not sure about, you know, we've got a record where we report any action that we've done. Like if someone comes in and starts causing some things, some issues,  or they start saying some weird things, I can be like `Okay, that name sounds familiar'. I can go search in our Discord and be like, `Oh they've caused problems before'. Then, I'll know to go after it. (P31, F, 34)
\end{quote}

\paragraph{Coordinate with Active Online Mods for Task Support and Transition}

During the streaming events, Some mods were active in the live streaming chatroom; some were active in both the live streaming chatroom and the Discord server; some might only be online in the Discord server without active engagement. These only online in Discord were on call anytime. Mods working on moderation tasks would ask other active mods for support if necessary and coordinate tasks with other mods before leaving.

Mods sometimes had to seek support due to the increased workload or the shortage of active mods in the chat. P7 (M, 19) said, \textit{``It's more like the streamer goes live, and there will probably be some mods that go and watch him as well. We don't have schedules, like you mod now, and now you have to mod. It's more like okay, he's live, and if it's really necessary, people can tag all of the moderators in the Discord and tell them if you can come help out.''}  According to P7, the mods usually had no schedule for the stream, but they were all online in Discord. They were on call to be active mods if others asked for support. For example, P36 (F, 20) said sometimes there was only one active mod in the chat, and the others were inactive, so he typed in the discord channel to ask other mods to take over while he did something else. 

In some cases, the stream experienced a sudden viewer increase, called a \textit{``raid''},  increasing the mods workload, so that they would like to ask other mods for help: \textit{``If a particular stream is having trouble or has extra viewers that they're not used to, it has a big raid or something, they might post in the mod channel `hey, can anyone come help out?' ''} (P5, M, 41).

Mods usually took responsibility and did whatever they could to help the community. However, they also needed breaks or do something for offline lives. In these cases, they would coordinate tasks with other mods to ensure the tasks were transferred. 
P4 (M, 35) stated that he usually took points and looked at the chat all the time when he was active; other mods would glance at the chat every so often; if he had to leave, he would ask other mods to replace his role. Mods would back each other up, like P27 (M, 21) said, \textit{``Sometimes it's just like, hey, is anyone here? I need to go make food,  and then we'll reply like, yeah, I'm here.''} 

As a head mod, the responsibility was larger; if they had to leave, they needed to ensure that the team worked well. P7  (M, 19) was the head mod for the channel and the discord admin as well: \textit{``I made the discord group, and it's more like once I made that, I always keep in an eye on it and I make sure everything is in order and that nothing is set up incorrectly, but at this point like I could just go off for months and it's not like the whole place would burn down. I also appointed another mod as discord admin who also has like extra permissions, so in case I can't  jump in, he can just do what I would do.''} According to P7,  the head mods assigned other mods as admins to gain \textit{``extra permission''} before leaving.

\paragraph{Discuss with Other Mods and the Streamer to Collaboratively Make Live Group Decisions}

To ensure moderation democracy and not make decisions by one single mod, mods would like to involve other mods to discuss what they should do and achieve a consensus. P15 (M, n/a) said,

\begin{quote}
A lot of things I usually end up asking are if we're going to make a severe change, like if the entire chat is spamming something like a really long copypaste. Usually, whenever I'm in a voice chat, I'll ask them, ``how long do you want to let this run?'' And instead of me being a dictator or letting someone be a dictator... I like to try and let other people be involved with the decision-making process where we all discuss how long we're going to let this run for and we come to a clear concise consensus to where this ends here. 
\end{quote}

According to P15, mods avoided being \textit{``dictators''} and discussed with other mods to \textit{``come to a clear concise consensus''}. Similarly, P11 (M, 23) would like to ask other mods to \textit{``mutually agree on what to do with the person.''} Usually, individual mods felt it difficult to make the final decision and needed a \textit{``second opinion''} (P32, F, 19). P29 (M, 29) elaborated, \textit{``We don't usually have any direct instruction from the streamer... We kind of work stuff through on ourselves. We have a Discord. We all kind of collaborate, and if we feel differently about things that we can talk stuff through.''}  In rare cases with severe issues, the streamer was also involved. For example, P10 (Trans Male, 45) stated, \textit{``If there are questionable situations, we'll have discussions amongst the mods or with the streamers about what to do in certain situations. In niche cases where we don't know about this, we have a discussion about it on Discord or in a private message about what guidelines we want to have.''} In this case, there were no clear standards for all situations, and mods had to discuss to decide the \textit{``guidelines''}.

\subsubsection{Foster Relationship both Off-stream and In-stream to Reinforce Task Collaboration}
In the stream, mods got to know each other and built relationships within the team. We clarified that mods also built no-task-oriented relationships; here, we focused on how they fostered relationships to reinforce task collaboration. Off-stream interaction such as sharing interests and playing video games can also help build friendships. Many mods would like to consider these with frequent interactions on Discord or meeting offline as friends who shared similar values and interests; at the same time, they might consider these with little or no communication as colleagues. P24 (M, 23) said, \textit{``I've made friends with a few of them. I wouldn't say that I'm friends with every moderator in every stream, but at least the ones I did talk to we're on good terms.''} Sometimes, they tried to be friends but were not sure about other mods' thoughts. P25 (M, 18) stated, \textit{``I feel like we've become friends over time, but again, everybody's hidden behind the username, and you can't really become friends. But I feel like for a long period of time you can get to know somebody, but you don't know their true self until you meet them in person.''} In this case, though mods considered these with \textit{``a long period of time''} communication friends but still doubted their \textit{``true self''}. Mods also tried to build friendships with others and believed that communication was important because the relationships can facilitate task collaboration and avoid team conflicts. Good relationships can make you \textit{``get along a lot easier''} with other mods (P28, F, 18).

\begin{quote}
Mod: Now, I try to get along with all the mods because I think…I haven't experienced this myself, but I've seen it; mod in-fighting is a really bad thing for the stream, the streamer, the mods, you know, that's just a bad thing all around for everybody. 

Interviewer: What is mod in-fighting?

Mod: When mods don't agree with each other, and they let their egos get in the way of doing their job, and it just causes problems with other mods, and they just argue.'' (P19, M, 21)
\end{quote}

According to P19, his experience to see mod in-fighting made him try to build good relationships with the team.   

\subsection{Ways to Contribute to Moderation Standards}
The second research question inquired how mods contributed  to the moderation standards in the team. Each community on Twitch has different standards. Mods mainly contributed to moderation standards in two ways: (1) collaboratively working with the streamer on the explicit community rulebook, including chat rules, channel rules, and completed list of rules and records only visible to the moderation team, and (2) individually configuring the implicit norms and criteria if there was no rulebook. We clarified that rulebook and norms were not exclusive; sometimes, though mods had the rulebook, they still used implicit norms to deal with specific situations.  

\subsubsection{Collaboratively  Work with the Streamer on the Community Rulebook}
Mods contributed to the community rulebook in two ways: assisting the streamer to polish/update the rulebook through discussion with the streamer if the streamer initialized it, or working with the streamer or the head mod to establish the rulebook if the community did not have one.

\paragraph{Assist the Streamer to Polish the Rulebook}
14 mods explicitly said that the streamer set up the guidelines for their channels, and many of them stated that the mods worked together with the streamer to polish the rulebook, though streamers made the final decisions. This process involved both the streamer's individual work and the mods-streamer collaboration. P13 (M, 20) reported that the initial set of rules was usually basic; over time, various things had cropped up; mods had to adjust the rules as needed. \textit{``The initial one was him, and then as a group. I think our streamer gets the final say on what he wants his rules to be, and we just enforce them. But he does take feedback very well,''} P31 (F, 34) added.  

Mods' contributions might be different based on streamers' experiences and preferences. Less experienced streamers would rely more on mods' feedback; more experienced streamers might dominate the development of the rulebook but also consider mods' opinions.  

\begin{quote}
	After they've brought up their rules, they're like, this is what I would like to see. But they don't have the experience with chat. And if they're a new stream or whatever, then they'll come up with their rules, and then they'll ask their moderators. What do you think about these? In your experience? Do these rules work? Is there anything you'd like to change? So ultimately, the rules are decided by the streamer and then the moderators, then either one to uphold those rules. (P23, M, 31)
\end{quote}

In this case, if the streamer had no moderation experience with chat messages or started a new stream without a clear idea about moderation in those categories, mods supported the streamer to uphold and modify the rules as needed. Other streamers might not take much input from mods: \textit{``The rules, the streamer often does, but we, the moderators do have a lot of input, well I wouldn't say a lot, but we do have a bit of an opinion as to what should count as a rule or not''} (P38, M, 18).

Mods' experience and their relationship with the streamer made their contribution to the rulebook differently.  

\begin{quote}
	The streamer sits down with the most trusted admins of the moderators. They're more referred to as the admins because, you know, they're trusted people, so there's a list of three people for our specific streamer who he goes to every time to figure out what to add, what to change, what to do. Then, it filters out to the rest of the moderators what's happening, and kind of, we ask their opinion second, so it ends up going around like everyone gets to put their opinion in, but the main criteria are decided by the streamer himself and three individuals that he trusts a lot. (P28, F, 18)
\end{quote}

According to P28, experienced mods can be considered trusted mods or admins; these mods contributed to the rulebook more significantly than other mods. The streamer mostly discussed with them and informed other mods.

\paragraph{Work with the Streamer or Head Mod with the Support of the Streamer to Establish the Rulebook}

Some channels did not have a  rulebook. Mods would like to try to proactively influence the community using their experts. P27 (M, 21) shared his experience about how to help a new streamer establish rules: \textit{``If I'm helping a new streamer set up, I will ask them like, Hey, what do you want this community to like what? Like, give me an overall view of how you want the community, and then from there, we derive some rules from it.''} In this case, the mods discussed with the streamer to have an \textit{``overall view''} and then derived some rules together. Mods also suggested the streamer with no rulebook borrow rules from other channels that usually had a complete list of rules. 

Some micro-communities had a clear hierarchy -- the streamer, the head mod, other mods. Head mods usually engaged in the communities for a long time to know the streamers' preferences; they led the rulebook development and took care of other mods.

\begin{quote}
	When I got modded in the beginning, there weren't really guidelines or anything. It was just like other mods told you, oh yeah, you should do this stuff, but that was more like tips. There weren't any like set guidelines. Now that I have sort of taken on the role of the head mod, and I have made comprehensive guides and a channel full of info. You know, stuff that might be not very clear to moderators so they can fall back on that if they have any questions. (P7, M, 19)
\end{quote}

\begin{quote}
    It always comes from the streamer, but some streamers have a head mod. For example, like a mod that takes care of all the mods, so if the mods have questions, they can just hit up that one person instead of always asking the question to the streamer directly. Um, but it's usually what the streamer wants. And yeah, I know some big streamers, like, for example, I have a friend who's a head mod for this streamer... I know he's the one who decides of the rules and everything, like he knows the streamer and he knows what he likes. So I know he's the one who takes care of all of that. (P36, F, 20)
\end{quote}
In these cases, Head mods helped the streamer to develop rules from the ground. Thus, the head mods developed most of the rulebook with either the streamer's or other mods' inputs. The benefits of establishing rulebooks were very obvious. Mods avoided using subjectivity to make decisions and take actions (P2, M, n/a).

\subsubsection{Work Individually with Limited Collaboration to Shape Community Norms}
If streamers didn't have a rulebook or didn't want to, mods figured out the moderation standards in various ways. We pointed out that most of these methods involved more individual work unless it is necessary to involve other mods or the streamer.  

\paragraph{Streamer's Expectation and  Mods' Experience and Judgment as Implicit Guidelines}
Mods used streamers' expectations and preferences as implicit guidelines. Obviously, the preferences can be shown in the explicit rulebook. In case of no rulebook, mods figured out streamers' expectations via directly talking with the streamers: \textit{``I usually try to get the streamer to sit down with me and to say what is the expected outcome of your community''}(P24, M, 23). Mods also stayed in the community for a long time, such as P19 (M, 31): \textit{``I already knew what he expected, just from conversations we've had, you know, not about modding. I watched his stream for many, many years and I knew basically just what is expected.''} According to P19, One way to figure out the streamer's expectations was to stay in the community. It can also help mods learn the community norms. 

\begin{quote}
	That's through understanding, like being part of that community. Because you've been a part of the community, you've been interacting in a way; you will already understand what is acceptable and what isn't acceptable and the usual guidelines for how things are dealt with. The streamer in question usually has a particular way that they deal with things, and that's been evoked many times, so that's how you deal with it in the future. (P9, M, 40)
\end{quote}

According to P9, mods understood streamers' expectations by watching \textit{``evoked''} violations and punishments (norms about how other mods collaborated),  also by interacting with other users to understand \textit{``what is acceptable and what isn't acceptable''} (norms about how users behaved).

In case streamers with no clear expectation or definition about rulebook, streamers either trust mods' judgment or empower mods to moderate with their experiences, as P20 (M, 43) said, \textit{``A lot of times, they don't even tell you. they trust you to use your judgment, just being basic human beings and knowing what's right what's wrong.''} 

The personal experience was about not only engaging in online communities but also reflecting on real-life scenarios. P01 (M, 18) described his judgment of the appropriateness of the content by imagining that these viewers were talking face to face: \textit{``One thing that I kind of really focus on also is `would you say this to a person directly face to face with that person?' You know what I mean. If you met this person uh... in real life, would you say that to their face? If you wouldn't actually say that to that person, then you will probably get banned for it.''} Mos also relied on working experience and educational background as references for their decisions. P04 (M, 35) stated, \textit{``I have a bachelor's in law, and that certainly helps me deal with people, and I also helped a little kid for a year, and also it's helpful.''}

As the implicit guidelines are vague and can cause mods to over-moderate in some situations, some mods said that streamers would correct them if their decisions were not in line with streamers' expectations. For example, P22 (M, 24) said he had no rulebook to follow and often used his common judgment: \textit{``If it's wrong in your eyes and the streamer doesn't think it's wrong, he's going to let you know.''} Similarly, P12 (F, 27) said she never communicated with the streamer to discuss their expectation and usually used her judgment: \textit{``Generally, I can apply my own rules to it, and if they don't like it, he will actually say `this is how I want timeout or ban to happen.' ''} Accordingly, streamers always had the final say about moderation decisions and could rectify the moderation action if necessary.

\paragraph{Maintain Standards with Other Similar Communities}
Some mods referred to similar communities with either clear rules or observed community norms. It can happen either within or outside the platform.   
P1 (M, 18) said, \textit{``If you can see this in other streamer's chat and get banned for, then you can get banned  in this chat.''}  In this case,  the mods also watched other live streaming channels; they might be viewers for other channels and saw violations and punishments in these communities and applied the norms to their moderated communities.  

Some mods specifically mentioned that they relied on rules on external platforms, which had more strict and complete rules. For example, P8 (F, 40)  gave an example of one community that he moderated: \textit{``Heroes Hype has very strict community guidelines and whatever because they are associated directly with Blizzard. So we kind of have to maintain the same community standards that Blizzard has on their own channel.''} Similarly,  P11 (M, 23) stated, \textit{``The Coalition, which is the studio who makes Gears of War, and Microsoft, came up with stream rules. So, if something goes against those rules, then we have to purge them or ban the person who said it.''} In these cases, the games might have official streaming accounts on Twitch, the gaming companies might apply developed rules for gaming to live streaming, or the mods had to maintain the same standards. 

\paragraph{Stick to the High-level Terms of Service}
Twitch keeps updating the community guidelines, which are complete than the channel rules. Some mods preferred to stick to the complete community-wide guidelines instead of the channel-specific rules. For example, 
\begin{quote}
	For the channel like the specific guidelines for the channel, we don't really do it much because we can mistake the specific guideline we have. We can't just go with that; you know what I mean. If anything needs to be added in the future, we will add and notify the mods about why this is being added or why we are doing whatever it is. But Twitch's guideline as a whole is a really interesting subject right now because they are changing their entire terms of service. (P1, M, 18)
\end{quote}

P1 preferred the terms of service of Twitch as guidelines because it was comprehensive. Additionally, P2 (M, n/a) stated, \textit{``Twitch has changed their terms of service because there had been a lot of complaints that rules weren't being clear enough and that Twitch admins were doing pretty much whatever they wanted.''} P1 pointed out the limitation of channel rules since they were developed by the moderation team and needed to be updated all the time but could still fall into mistakes. P2 pointed out the superiority of the updated terms of service. The high-level terms of service were complete, transparent, and developed by the platform; consequently, it was safe and easy to follow. In this case, there was not much communication and collaboration in the moderation team. 

\section{Discussion}

This work supplements prior work about volunteer moderation and explores how mods and the streamer work as a team to work on moderation standards and tasks. We identified three high-level themes about how the moderation team collaborates on tasks: off-streaming coordination and preparation (recruiting, notifying, and updating), in-stream real-time collaboration (monitoring and informing the team regarding potential incidents, coordinating for task support and transition, and discussing with the team to make live group decisions), and relationship building both off-stream and in-stream to reinforce task collaboration. We also found that mods contribute to moderation standards in mainly two ways: collaboratively working on the community rulebook with the streamer and other mods (assisting the streamer in polishing or working with the streamer/head mod to develop the rulebook), and individually working the community norms if there is not clear community rulebook (streamers' expectation and mods' judgment, standards from similar communities, and high-level terms of service). We clarify that this work mainly contributes to content moderation and live streaming context, and we use remote collaboration to explain these phenomena. In this section, we first discuss the nuanced differences that contribute to volunteer moderation; then, we discuss how the streamer and mods work as an effective team to some extent, using Olson and Olson's four components \cite{Olson2000DistanceMatters}. 
 
\subsection{Moderation Standards, Coordination Mechanism, Mods' Roles and Responsibilities}

\subsubsection{Multi-layered Standards Both Visible and Invisible to the Public}

Rules and norms are important to volunteer moderation. We found that the moderation team uses a mix of collaborative and individual processes to work on moderation standards development. A lot of work has disused the transparency and clarity of rules and norms (e.g., \cite{Cai2021UnderstandingTwitch,Gilbert2020iR/AskHistorians,Jhaver2019DoesReddit,Jhaver2019DidReddit}). Though users complain about the opaque nature of content moderation and many scholars argue for transparency, there is still something invisible to the public. The rulebook includes the visible chat and channel rules to the public and a list of details and shared documents to handle specific situations that are only visible to the moderation team. While the visible rulebook helps regulate viewers' behaviors and boost the community atmosphere, the rulebook for the team's internal reference might play a prominent role for mods to do tasks.  

Mods shape community norms by learning the moderation team norms and engaging with other viewers to learn the chat norms. These are not explicitly written but understood by the mods and shared in the team. Mods also refer to other similar communities that have clear rules. In this sense, we provide a nuanced understanding of the moderation standards and how mods work for them. Future work can explore the mechanism behind multi-layered standards like how they influence the platform governance, particularly how the social governance and algorithmic governance mechanisms influence each other \cite{Muller-Birn2013Work-to-rule:Wikipedia}. 

\subsubsection{The informality of Voluntary Work Dilutes Mods' Roles and Responsibilities}
Prior work suggests that an effective virtual team must have both implementer and completer-finisher but might not need the project coordinators \cite{Eubanks2016TeamFulfillment}. In the volunteer moderation context, the streamer performs the leading role of coordinator during the off-stream coordination and rulebook development, thus can be the completer to aggregate or update the rulebook development. In the stream, mods all become implementers; there is no specific completer-finisher; the streamer mainly focuses on streaming instead of moderating. Accordingly, streamers as coordinators play significant roles in ensuring enough active mods online, which is prominent in live streaming communities. Prior work also suggests that team leaders threaten team survival if they are too controlling \cite{Kraut2014TheGroups}. Similarly, we found that streamers empower mods to do the moderation (e.g., trust mod's judgment) as a way to build the team.  

Mods' roles and responsibilities can also weigh differently. Off the stream, the streamer-mod collaboration is the prominent collaboration;  in the stream,  the moderation process involves more mods collaboration. Prior work has classified many different roles mods play to help the streamer or the community \cite{Wohn2019VolunteerExperience,Matias2019TheOnline, Seering2022MetaphorsModeration}. In the moderation team, we found that there is no specific role regarding the task collaboration except that the head mods assign and notify things. Many mods state that being active online is important and that the task is more like \textit{``first come first served''} without fixed roles and responsibilities. Their roles seems contingent and fluid \cite{Seering2022MetaphorsModeration}.

The informality of the volunteer moderation might explicate this phenomenon, differentiating it from other distributed teamwork. Informal work includes all activities leading to the production of services for others outside a legal framework and can be community-oriented and based on mutual help and moral obligation \cite{Pfau-Effinger2010FormalEurope}. Mods as individuals consider moderation an informal work without the imposed regularity. They are not paid laborers and have no forced schedule to work with the streamer. Some mods with strong moral obligations notify each other and coordinate the availability; others describe that no show-up is also acceptable. Additionally, some mods might do more based on their availability, but there is no fairness about the workload (who should do what at what time).   

\subsubsection{Synchronicity Facilitates Explicit and  Implicit Coordination}
Most of the task coordination identified can be considered explicit coordination. The moderation team uses explicit coordination off and in the stream to ensure active online mods (e.g., recruiting new mods) and seek task support (e.g., asking for support). Mods also involve much implicit coordination when multiple mods are active and dealing with violators in real-time. The implicit coordination is highly interdependent, time-sensitive, stressful because of the synchronous interaction in the chat.  

The implicit coordination derives from the shared mental model \cite{Rico2008TeamApproach}. Shared mental models can be developed by staying in the community for a long time and also by using guidelines and policies \cite{Kittur2010BeyondGroups}. In this sense, mod's contribution to moderation standards is a good case of shared mental model development, actively engaging in the rulebook development and also community norms development via staying in the community for a long time. Thus, they mutually understand the community standards and what they should do. For example, if they see too many mods in the chat, some will self-deactivate and work on personal things, or only keep a glance of the chat without intervention unless some trouble happens or workload increases. Mods anticipate what other mods are likely to do and adapt their behaviors to facilitate the moderation tasks without explicitly discussing who should do what.  

Though sometimes there is a task conflict, they trust each other and don't see it as a big problem. The affordances of live streaming moderation bring all the active mods online at the same time with the same content \cite{Hamilton2014StreamingMedia}. Though multiple mods notice a violation and moderate it simultaneously (e.g., timeout twice), they can easily see it almost at the same time and consequently revoke the moderation action. Some mods can also have brief real-time conversations to resolve the conflict. To some extent, the synchronicity of live streaming facilitates implicit coordination and makes moderation function smoothly without clear roles and responsibilities.

\subsection{The Streamer and Volunteer Mods Form an Effective Team}
 Olson and Olson's four-component framework \cite{Olson2000DistanceMatters} is usually applied in the professional and formal context. Thinking about how our results may align with the framework, we found that this framework can potentially be applied to this voluntary and informal context to show how volunteer mods and the streamer can work effectively as a team. The collaboration and technology are ready; the common ground is clearly built; the dependency of work is varied but mostly loose; thus, the remote collaboration is generally effective and to some extent productive, though mods experience some challenges. 

\subsubsection{Common Ground}
Prior work suggests that sharing knowledge and understanding the awareness of the state of other team members is an essential part of common ground, and working for a long time on a certain project can build common ground \cite{Olson2000DistanceMatters, Bjrn2014DoesCollaboration}. Olson and Olson pointed out that people who have established a lot of common ground can communicate well even over impoverished media \cite{Olson2000DistanceMatters}. In that sense, mods' contribution to moderation standards involves the common ground building: collaboratively working with a streamer/head mod to develop the rulebook and sharing it with the team, individually working on community norms such as staying in the micro-community for a long time to know streamer's expectation or to learn by observing. Task collaboration process also contributes to common ground buildings, for example,  off-stream coordination such as updating information and notifying each other, and in-stream communication. Building good relationships with other mods is another way to build common ground and, in turn, can facilitate communication and avoid intragroup conflict. 

\subsubsection{Dependency/Coupling of Work}
Coupling of work refers to \textit{``the extent and kind of communication required by the work''} \cite{Olson2000DistanceMatters}. Work with strong dependency is usually non-routine and ambiguous, requiring the collective work of team members engaging in frequent and complex communication with short feedback loops, and vice versa \cite{Olson2000DistanceMatters}. In the volunteer moderation team, we found that mods' tasks involve both tight and loose dependency of work. In communities with clear moderation standards, the common ground about the violations and punishment is clear, so the moderation task is procedural and straightforward. In this sense, the dependency of work is comparatively loose. In communities with no clear moderation standards (e.g., mods' own judgment), or scenarios that can't be covered by these standards, the ambiguity of the tasks requires communication to clarify. In this sense, the dependency of work is comparatively tight. Most of the time, moderation tasks are routine tasks that any active mods can do it. In the organizational context, the tightly coupled work is challenging to do remotely, and technology does not support the rapid back and forth in conversation or awareness and repair of ambiguity \cite{Olson2000DistanceMatters}. However, we did not see a clear challenge or difficulty regarding the technology usage, in line with prior work \cite{Bjrn2014DoesCollaboration}. We explain this in technology readiness.  

\subsubsection{Collaboration Readiness}
Since most volunteer mods are selected from active viewers and viewers having a close relationship with the streamer \cite{Wohn2019VolunteerExperience}, either having similar value with the streamer or having a positive influence on the community value. Their background and experience are homogeneous and consistent; there is less barrier to collaborate remotely from the beginning \cite{Bjrn2014DoesCollaboration}. The nature of moderation work is online and remote and has a culture of sharing and collaboration. Mods assume to work with others remotely. Thus, mods are appropriately ready to collaborate remotely.  

\subsubsection{Technology Readiness}
The current technologies provide multi-modal interaction in the team. Mods use different technologies, including the Whisper function within the Twitch platform and third-party tools like Discord and Skype. While mods can work as a small group (2-3) via Whisper function to have a group chat, most communication is parallel and happens on Discord, offering both voice chat and text chat. The text chat is mostly served as asynchronous communication supporting time-delayed interaction and provides anytime/anywhere flexibility, while the voice chat is mainly served as synchronous communication with real-time interaction, allowing immediate discussion and clarification \cite{Sere2011OnlinePerformance}. Additionally, the streamer can directly answer mods' questions or remedy mods' actions in the chat room as mods are doing the moderation tasks. Both live streaming and third-party platforms cover video, audio, and textual communication. Mods did not complain much about the technological infrastructure for remote collaboration. 

Twitch provides various moderation tools for the team to do tasks \cite{Cai2019CategorizingTwitch}, but our results reveal that most of the communication happens on Discord. It seems like the team neither uses moderation tools to communicate nor uses communication tools to do tasks. There should have an artifact that supports both tasks and communication. However, we don't know whether the team purposely separates them on different platforms or they are forced to do so because of limited choices, requiring further investigation.  

\subsection{Design Implications}

Though mods do not complain about the technology usage, they do apply various available tools that are not deliberately designed for moderation teamwork. They also face challenges to ensure a number of active mods online when the streamer goes live and to consistently enforce moderation standards that are vague and varied, to some extent. We provide several suggestions to simplify the tools they have used to switch back and forth and reduce the cost of collaboration. These features are necessary to moderation teamwork and should be integrated into the ecosystem of the live streaming platform by either Twitch or third-party developers.

\subsubsection{Coordination Support Design}

\paragraph{Tools to Coordinate Team Availability}
The current Twitch platform only allows streamers to post their schedules to the public but lacks a function to coordinate them with mods' schedules. We found that schedule coordination involves much back-and-forth communication on the third-party platform. Thus, we suggest a collaborative calendar-like design to automatically coordinate schedules for mods and streamers. For example, mods can also add their availability to the streamer's schedule and indicate their availability level (High, medium, low). The design can be embedded into either the live streaming platform or the moderation tools.

\paragraph{Tools to Signal Attention and Roles}
As we discussed, there is no specific role in moderating the chat, and only ensuring active mods can sometimes engender fluid roles and task conflicts. Mods neither know other mods' preferences nor have a clear idea of other mods' roles and might have to figure out who is doing what. We recommend an attention indicator design with social signals \cite{Im2020SynthesizedHistories} to show mods' anticipated roles in the chatroom, such as bot manager, socializer, punisher, rule poster. For example, Twitch has a rich badge system to indicate viewers' contribution and participation; it may add another layer of badges to show mod's roles in the moderation team only visible from a mod's view. These role indicators can also give the streamer an overview in advance and coordinate mods with specific roles they need.

\subsubsection{Rulebook Development Design}
\paragraph{Tools to Support Visible Rulebook Development}
Twitch has a community guideline targeting general users with easy-to-read terms. This guideline also suffers from vague definitions for some terms and does not cover specific needs for some micro-communities. We suggest a customized mechanism that can aggregate chat and channel rules based on streamers' (e.g., female or marginalized streamers with specific needs \cite{Uttarapong2021HarassmentNegativity}) and communities' characteristics (e.g., same game category, similar topics) to provide templates for new streamers or streamers starting new topics/categories. These communities might suffer from similar harmful content and inherit similar values and visions. The aggregate rules can clearly show what is allowed or not in the communities, and streamers can also tailor the generated template.   Additionally, a mechanism to evaluate the rule efficacy by comparing the existing rules with blocked content to help mods differentiate \textit{``good''} or \textit{``bad''} rules, and as a result, to update the rulebook \cite{Vaidya2021ConceptualizingModeration}, can also be considered.  

\paragraph{Tools to Support Invisible Rulebook Development}
A clear rulebook helps mods reduce the subjectivity of the decision-making process. The invisible rulebook is critical for mods to make the appropriate decision. Mods currently use other artifacts to develop the invisible rulebook and share them on third-party platforms. To better support the teamwork, we suggest an artifact that can facilitate the invisible rulebook development, for example, a mechanism to summarize violations and automatically organize the list of repeated violators. The streamer or head mods can give its access to the moderation team and allow all mods to update and manage.  

\subsubsection{Relationship Building Design}

Communication and coordination might facilitate relationship building and in turn, build common ground. While mods have different expectations regarding their relationship with the streamer and viewers \cite{Wohn2019VolunteerExperience}, we found that mods try to foster a relationship with other mods but only consider frequently communicated mods friends. As volunteer mods, the perceived relationship might directly influence their continuous engagement in the community. For example, conflicts about what should be labeled as \textit{``bad''} behaviors in the moderation team can cause mod resignations and status removal \cite{McGillicuddy2016ControllingWork}. Thus, design to support relationship building and avoid team conflict is also critical.

We recommend mechanisms to facilitate informal communication and collaboration,  similar to the community-based groupware supporting lightweight communication to allow easy initiation of interaction and an overview representation of the community that shows who is available and what individual task they are working on \cite{Gutwin2008SupportingGroupware}. For example, in Discord mod-only channels, by adding labels about social topics, mods can choose labels to show their interest and intention, to facilitate mods to form sub-groups and communicate more. The initialized communication at an early stage can also increase their commitment through its influence on the group atmosphere \cite{Dabbish2012CommunicationTeam}.

\subsection{Limitations}
This study also has several limitations. First, we only interviewed mods to explore teamwork, but we found streamers as team leaders also played significant roles. Future work can explore streamers' perspectives about teamwork to supplement our research. Second, we only focused on a single live streaming platform --- Twitch, which has some unique affordances. We do not know how well these findings can be generalized to other live streaming platforms or even asynchronous communities applying volunteer moderation. Third, our results indicate that community size might be an important factor for teamwork, but we cannot quantify how community size influences the effectiveness of teamwork. It seems larger communities with lots of mods need more well-organized teamwork. Future work can explore the relationship between the threshold of community size and team evolution dynamics. Lastly, we do not consider mods' individual characteristics in the analysis. According to our findings, tenure might lead mods to senior/trusted mods and rely more on implicit coordination in the collaborative process. Future work can integrate mods' individual characteristics and explore their impacts on moderation teamwork. 

\section{Conclusion}
In this paper, mods contributed to moderation standards in two ways and collaborated on tasks in three ways. We outlined how volunteer mods work as an effective team: their contribution to the moderation standards facilitates common ground building; the multi-platform,  multi-modal communication, and the nature of active community users engender collaboration readiness and technology readiness; and most routine tasks were loosely dependent. We also identified the multi-layered standards, the coordination mechanism, and mods' roles and responsibilities and discussed how synchronicity of communication and informality of volunteer moderation contributed to these phenomena. Designs to facilitate collaboration support, rulebook development, and relationship management are also suggested. 

\begin{acks}
This research was funded by National Science Foundation (Award No. 1928627). Thanks to the research assistants in SocialXLab at NJIT for data collection.
\end{acks}

\bibliographystyle{ACM-Reference-Format}
\bibliography{references}

\end{document}